# The origin of second-order transverse magnetic anisotropy in Mn12-*acetate*

A. Cornia

*Dipartimento di Chimica and INSTM, Università di Modena e Reggio Emilia, via G. Campi 183, I-41100 Modena, Italy*

R. Sessoli, L. Sorace, D. Gatteschi

*Dipartimento di Chimica and INSTM, Università di Firenze, Via della Lastruccia 3, 50019, Sesto Fiorentino , Italy*

A. L. Barra

*Laboratoire des Champs Magnetiques Intenses – CNRS, Grenoble Cedex 9, F-38042 France*

C. Daiguebonne

*Groupe de Recherche en Chimie et Métallurgie , INSA, 35043 Rennes Cedex, France*

**Abstract.** The problem of the role of transverse fields in Mn12-acetate, a molecular nanomagnet, is still open. We present structural evidences that the disorder of the acetic acid of crystallization indices sizeable distortion of the Mn(III) sites, giving rise to six different isomers, four of them with symmetry lower than tetragonal. Using a ligand field approach the effect of the structure modifications on the second order transverse magnetic anisotropy, forbidden in tetragonal symmetry, has been evaluated. The order of magnitude of the quadratic transverse anisotropies well agree with the values derived by the analysis of the field sweep dependence of the hysteresis loops performed by Mertes et al. ( Phys. Rev. Lett **87**, 227205 (2001)) and allows to better simulate the EPR spectra.

PACS numbers:75.45.+j, 61.10.Nz, 31.15.Ct

# I. INTRODUCTION

Mn12-*acetate* is a molecular compound of formula [Mn$_{12}$O$_{12}$(CH$_3$COO)$_{16}$(H$_2$O)$_4$]·2CH$_3$COOH·4H$_2$O which comprises clusters of twelve manganese(III-IV) ions, characterized by S$_4$ symmetry in the tetragonal lattice[1]. It has been recently the object of intense investigations for its unusual magnetic properties. Each cluster is in fact characterized by a *S* = 10 ground spin state experiencing a strong uniaxial anisotropy. Below 3 K the magnetization is frozen due to the energy barrier of *ca.* 60 K which hampers the free reorientation of the magnetization[2]. The *S* = 10 ground state of the cluster is split into 21 sublevels describing a double-well potential according to the first term in the hamiltonian

$$\mathbf{H} = D\mathbf{S}_z^2 - g_z\mu_B H_z \mathbf{S}_z \qquad (1)$$

which describes the axial anisotropy and Zeeman interaction with a magnetic field $H_z$ directed along *z*. The quantum nature of the system is at the origin of the observation of steps in the hysteresis cycle[3,4], which occur whenever levels on opposite sides of the barrier have the same energy and tunneling through the barrier can occur. However, in order to observe tunneling the hamiltonian must contain terms which do not commute with $\mathbf{S}_z$. A transverse magnetic field of any origin can induce tunneling but in zero or axially-applied external field the dipolar or hyperfine fields are too weak to justify the observed tunneling rate. Second- and fourth-order transverse anisotropies described by the hamiltonian:

$$\mathbf{H'} = E(\mathbf{S}_x^2 - \mathbf{S}_y^2) + B/2(\mathbf{S}_x^4 + \mathbf{S}_y^4) \qquad (2)$$

promote tunneling, the E term allowing transitions between states whose *M*, eigenvalue of $\mathbf{S}_z$, differ by multiples of two. However, the *E* parameter must vanish according to the fourfold crystallographic symmetry of the clusters. The symmetry allowed fourth order term in (2), permits tunneling every fourth step. However, all the transitions from M=-10 to M'=9,8,..etc., are experimentally observed on a similar footing. Several authors [5-7] have recently suggested the presence of symmetry breaking effects (i.e. a nonzero E parameter) in conjunction with a transverse magnetic field which activates odd transitions. To justify the observed behavior, the authors have considered the *E* parameter to vary locally in the crystal according to a broad gaussian distribution. Among the possible physical origins of local symmetry lowering Chudnovsky *et al.* have proposed crystal dislocations [8]. Even if we cannot exclude their presence in the crystals of Mn12-*acetate* we want to show in this Letter that most of the Mn12-*acetate* molecules in the crystal do not actually possess four-fold symmetry. In fact, the tetragonal symmetry is disrupted by the acetic acid molecules of crystallization, which are disordered over two sites around twofold axes. A detailed X-ray diffraction analysis at low-temperature has evidenced that the anisotropy axis of a manganese(III) site is significantly bent when hydrogen-bond interactions with the acetic acid occur. In principle, this gives rise to six different types of clusters, whose relative abundance can be calculated assuming statistical distribution, as will be shown below.

## II. CRYSTAL STRUCTURE ANALYSIS

Mn12-*acetate* was first synthesized and structurally characterized by room-temperature X-ray diffraction methods in 1980[1]. The [$Mn_{12}O_{12}(CH_3COO)_{16}(H_2O)_4$] clusters develop around $S_4$-axes of tetragonal space-group I-4 and are consequently crystallographically axial. The lattice also comprises four non-coordinated water molecules (O13) (see Figure 1 for the atom labeling scheme) and two acetic acid molecules of crystallization (O14, O15, C9, C10) per cluster. The latter are located around two-fold axes between adjacent Mn12 units and are consequently disordered over two equally-populated positions. Both the coordinated water molecule O12 and the oxygen atom of an acetate ligand in the cluster (O6) are found at H-bond distance from the acetic acid oxygens O14 and O15, respectively. However, the precise pattern of hydrogen-bond interactions has not been determined so far. Also, no explanation has been provided for the unusually large Debye-Waller parameters of C4 and O6 in the acetate ligand O5-O6-C3-C4 ($B_{eq}$ = 15.06 and 4.35 Å$^2$, respectively) as compared to corresponding atoms in the remaining acetate moieties (4.07 - 4.38 Å$^2$ for methyl carbon atoms and 2.19 – 3.25 Å$^2$ for oxygen atoms, respectively). In order to further investigate these issues, we collected a set of X-ray diffraction data on Mn12-*acetate* at 83 K. We summarize here the most relevant findings, while all details concerning the experimental setup and the crystallographic calculations will be reported elsewhere[9]. With our analysis, we have been able to determine the correct orientation of the acetic acid molecule, whose hydroxyl oxygen atom O15 forms a strong hydrogen bond with O6 [O15…O6 = 2.871(6) Å, O15-H1…O6 = 175.4(3)°], as depicted in Figures 1 and 2. More importantly, we have found that the displacement ellipsoids of O6, C4 and, to a lesser extent, of O7 and C3 are abnormally elongated, pointing to the presence of disorder effects on the acetate ligand too. An alternative structural model has then been used in which the acetate ligand is allowed to reside in two different positions (A and B). The best-fit populations are 0.46(1) and 0.54(1) for A and B, respectively. They strongly suggest that the disorder of the acetate ligand may result from the hydrogen-bond interaction with the disordered acetic acid molecule, which leads to a sizeable displacement of O6 toward O15 (Figure 2). In principle, each of the four symmetry-related acetate ligands derived from O5-O6-C3-C4 can be involved in H-bonding interactions of this type. Six different isomeric forms of Mn12-*acetate* can thus be envisaged which differ in the number (n = 0,1,…,4) and arrangement of hydrogen-bound acetate ligands (Figure 3). Clearly, strict axial symmetry can be retained only in the case of a regular pattern of n = 0 and n = 4 isomers. However, this would lead to a supercell with doubled lattice constants (a'=b' = 2a) for which no experimental evidence has been found. We conclude that the average molecular symmetry of Mn12 clusters is lower than axial. The relative abundance of the various species reported in Table 2 has been calculated assuming an equal probability of sites A and B, in agreement with the experimentally determined SOFs which are very close to ½.

### III. CALCULATION OF THE MAGNETIC ANISOTROPY

The most important contribution to the magnetic anisotropy in Mn12-*acetate* is expected to arise from the Jahn-Teller distorted manganese(III) sites, although different terms like Dzyaloshinski-Moryia interactions have been invoked [10]. We have shown in a previous publication[11] that the magnetic anisotropy of manganese(III) ions can be predicted with a great accuracy by taking into account the crystal field generated by the ligands and introducing the real geometry of the coordination sphere using the Angular Overlap Model[12]. This is a ligand-field model which uses molecular orbital oriented $e_\sigma$ and $e_\pi$ parameters for each donor atom, and is particularly well suited to account for angular distortions in the ligand field. The encouraging results of a first attempt of rationalization of the zero-field splitting (ZFS) of Mn12-*acetate* on the basis of AOM [13] have suggested that this approach might provide reasonable estimates of the magnetic anisotropy differences among the various isomers, eventually elucidating features of the low temperature spin dynamics.

Due to disorder effects, four different coordination geometries must be considered for the manganese(III) ions, namely two for Mn2 and two for Mn3 (see Figure 2). Manganese(IV) sites were neglected given the small expected contribution to the global ZFS. To avoid over-parameterization of the system, the interelectronic repulsion Racah parameters B and C and the spin-orbit coupling coefficient $\zeta$ have been set to the free ion values (B=1140 cm$^{-1}$, C=3675 cm$^{-1}$, $\zeta$=315 cm$^{-1}$), while the effect of covalence on spin-orbit coupling has been accounted for by setting the orbital reduction factor (k) to 0.75. The values of the $e_\sigma$ and $e_\pi$ parameters for each donor atom, taken from literature data, have been corrected for the actual metal-ligand distance assuming an exponential dependence. The angular coordinates are those obtained from the structure determination, while the ligand-field parameters are calculated as described elsewhere [12].

In Table I we report the results obtained from the calculation. As can be seen, the interaction with the acetic acid molecule induces only minor changes in the D and E parameters as well as in the direction of the easy axis.

In order to evaluate the departure of the overall anisotropy from tetragonal symmetry we have first evaluated how the single-ion contribution is projected on the $S = 10$ ground spin state. The projection coefficients strongly depend on the wavefunction describing the ground state. If the coupling scheme described by Villain et al.[14] is assumed the projection coefficient is identical for all the manganese(III) ions ($d_2=d_3= 0.02845$). The total anisotropy tensor for each of the six species shown in Figure 2 has been then calculated by performing the summation

$$\mathbf{D}_{tot} = d_2 \sum_{i=1}^{4} \mathbf{R}_i^T \mathbf{D}_2^{\alpha(i)} \mathbf{R}_i + d_3 \sum_{i=1}^{4} \mathbf{R}_i^T \mathbf{D}_3^{\alpha(i)} \mathbf{R}_i \qquad (3)$$

where $\alpha(i)$ = A or B. In the above Equation, $\mathbf{D}_2^{\alpha(i)}$ and $\mathbf{D}_3^{\alpha(i)}$ are the single-ion ZFS tensors for the Mn2 and Mn3 sites generated by the $i$-th symmetry operation of the $\mathbf{S}_4$ point-group.

The resulting $\mathbf{D}_{tot}$ tensor turns out to be axial and diagonal in the crystal axes reference frame only for $n = 0$ and $n = 4$. In the other four cases non-zero off diagonal terms are present and diagonalization of the matrices provided the D and E parameters along with the angle θ between the easy axis and the crystallographic $c$ axis (Table II). The agreement between the calculated D parameter and the experimental one found by EPR [15], Inelastic Neutron Scattering[16], and Torque Magnetometry[17], is acceptable considering the approximations involved (definition of the ground-state wavefunctions and the related projection coefficients, neglect of higher-order terms, etc.). The D parameter is equal for all the species within ±2% and the easy-axis direction does not deviate significantly from the crystallographic $c$ axis (<0.5°).

Another type of defect, namely the flipping of the Jahn-Teller elongation axis of Mn3, affects the magnetic anisotropy and accelerates the magnetic relaxation, as it has been clearly observed in other Mn12 clusters containing carboxylate ligands[18]. A faster-relaxing species is present in Mn12-*acetate* crystals as well, although in very low concentration so as to escape detection by X-ray diffraction methods[19]. We have evaluated the effect of this type of defect in Mn12-*acetate* by simply flipping the corresponding ligand field parameters values between the two different elongation axis of Mn3 as observed in [18], and assuming the same polar angles with respect to the "standard" site. The results for the overall anisotropy are also reported in Table II. It is quite evident that such a species presents a stronger transverse anisotropy as well as a much larger deviation of the easy axis from the crystallographic *c* axis (9 °).

## IV.    COMPARISON WITH EXPERIMENTS

To compare our results with EPR spectra we show in Figure 4 the high field region of the experimental polycrystalline powder spectra recorded at T= 35 K with the exciting frequency of 525 GHz [15]. The simulation has been obtained by using the second and fourth order axial parameters as in [16] with the distribution of second order transverse terms as indicated in Table II. Only B has been left free to vary and a good simulation has been obtained with B=4.0x$10^{-5}$ K, very close to that derived by INS experiments[16]. The agreement with the experimental data improves significantly compared to the previous simulation where only the fourth order transverse term was considered [15]. Spectra on a fresh single crystal have been recorded and even if a detailed analysis of these spectra will be presented elsewhere we show, in the inset of Figure 4, the spectrum obtained with the field along the *a* crystallographic axis at 5 K and 95 GHz. The spectrum clearly show that the lines are not simply broadened but at least two signals with comparable intensity can be detected. This confirms our hypothesis of a discrete number of isomers with slightly different parameters of spin-hamiltonian rather than a broad distribution of them.

The presence of a distribution of magnetic anisotropy in crystals of Mn12ac have also been experimentally evidenced by a detailed study of the dependence of the steps in the hysteresis loops on the field sweeping rate[5-7]. The ground state tunnel splitting $\Delta_{\pm 10}$ is practically unaffected by such small values of E but this is not the case for resonances occurring at higher fields. In order to compare our study with the experimental results in [5] we have evaluated the tunnel splitting induced by the presence of the second order magnetic anisotropy for the transition between the M=-10 and M'=4 state (N=6 transition). In Table II we report the calculated tunnel splitting for this transition by exact diagonalization of the hamiltonian matrix, assuming the other parameters of the spin hamiltonian as in [16]. The $\Delta_{-10,4}$ for such a transition is zero for species *n*=0 and *n*=4, in agreement with their higher symmetry, and varies between 7.0x$10^{-9}$ K and 2.0x$10^{-7}$ for the others. These values are in very good agreement with the distribution of tunnel splitting, centered around $\Delta_{-10,4} \approx 3 \times 10^{-8}$ K , used in [5] to scale the experimental data.

If our calculation is repeated by neglecting the fourth order transverse term the calculated splitting is several orders of magnitude smaller, confirming that this term cannot be neglected, as suggested by the simulation of the EPR spectra.

In conclusion we have provided an unambiguous experimental evidence that most of the molecules in Mn12-*acetate* experiences a symmetry lower than fourfold despite the tetragonal crystal symmetry. The order of magnitude of the quadratic anisotropy has been determined and found to be distributed around discrete values, which are in agreement with the distribution used to justify experimental results on the dynamics of the magnetization [5]. Our results show that the magnetocrystalline anisotropy is only slightly affected by perturbations of the crystal structure involving the metal coordination. It seems therefore quite unreasonable that dislocations, even if present, could induce substantial modification of the magnetic anisotropy at long distance as proposed in[8]. Dislocations can indeed be the source of a further broadening of the distribution of magnetic anisotropy, but the presence of disordered acetic acid remains the main source of the quadratic transverse anisotropy.

We are indebted with T. Roisnel for his assistance in the low-temperature X-ray data collection. The financial support of Italian MIUR and CNR and of the EC through the TMR program MOLNANOMAG (n°.HPRN-CT-1999-0012) is acknowledged.

**Table I.** Calculated Mn(III) magnetic anisotropy parameters.

| Site | D (K) | E (K) | $\delta$ (°)[a] |
|---|---|---|---|
| Mn2A | -4.92 | 0.40 | 11.6 |
| Mn2B | -5.27 | 0.27 | 10.7 |
| Mn3A | -4.57 | 0.10 | 37.2 |
| Mn3B | -4.40 | 0.07 | 37.1 |
| Mn3 flipped | -4.64 | 0.06 | 58.4 |

a) $\delta$ is the angle between the easy-axis direction of each manganese site and the crystallographic *c* axis.

**Table II.** Calculated magnetic anisotropy and tunnel splitting of the isomers of Mn12-*acetate*.

| isomer | concentration | D (K) | E(K) | $\theta$ (°) | $\Delta_{-10,4}$ (N=6)[a] |
|---|---|---|---|---|---|
| n=0 | 6.25% | 0.759 | 0 | 0 | 0 |
| n=4 | 6.25% | 0.797 | 0 | 0 | 0 |
| n=1 | 25% | 0.769 | 2.34x10$^{-3}$ | 0.3 | 8.8x10$^{-8}$ |
| n=2 cis | 25% | 0.778 | 1.87 x10$^{-4}$ | 0.4 | 7.0x10$^{-9}$ |
| n=2 trans | 12.5% | 0.778 | 4.70x10$^{-3}$ | 0 | 1.7x10$^{-7}$ |
| n=3 | 25% | 0.788 | 2.35x10$^{-3}$ | 0.3 | 8.9x10$^{-8}$ |
| flipped | undeterm. | 0.754 | 4.39 x10$^{-2}$ | 9.0 | 7.3x10$^{-7}$ |

a) The spin hamiltonian parameters are the same as in [16] where the E term reported in the 4[th] column has been added.

**Captions to Figures**

**FIG. 1**. Structure of the $[Mn_{12}O_{12}(CH3COO)_{16}(H_2O)_4]$ cluster at 83 K viewed slightly off the $S_4$ axis, together with the lattice water (O13) and disordered acetic acid (O14, O15, C9, C10) molecules (the two equivalent positions are differentiated by solid and empty bonds ). Hydrogen atoms on O12 and O15 are depicted as small spheres, while the remaining hydrogen atoms are omitted for clarity. The network of hydrogen bonds is shown by dashed lines.

**FIG. 2**. Coordination sphere of Mn2 and Mn3, as determined from anisotropic refinement of the displacement factors of C3, C4, O6 and O7 (left) and from an isotropic model with disorder fitting (right). Thermal ellipsoids are at 50%-probability level. Methyl hydrogen atoms are omitted for clarity.

**FIG. 3**. The six hydrogen-bond isomers of Mn12-*acetate* as discussed in the text.

**FIG. 4**. High Field EPR spectra of a polycristalline sample of Mn12-*acetate* at 35 K and 525 GHz (bold line) , simulation obtained in [15] (line & crosses), and with the inclusion of a distribution of second order anisotropy terms according to Table II (solid line). In the inset the single crystal spectrum recorded at 5K and 95 GHz with the magnetic field parallel to the *a* axis.

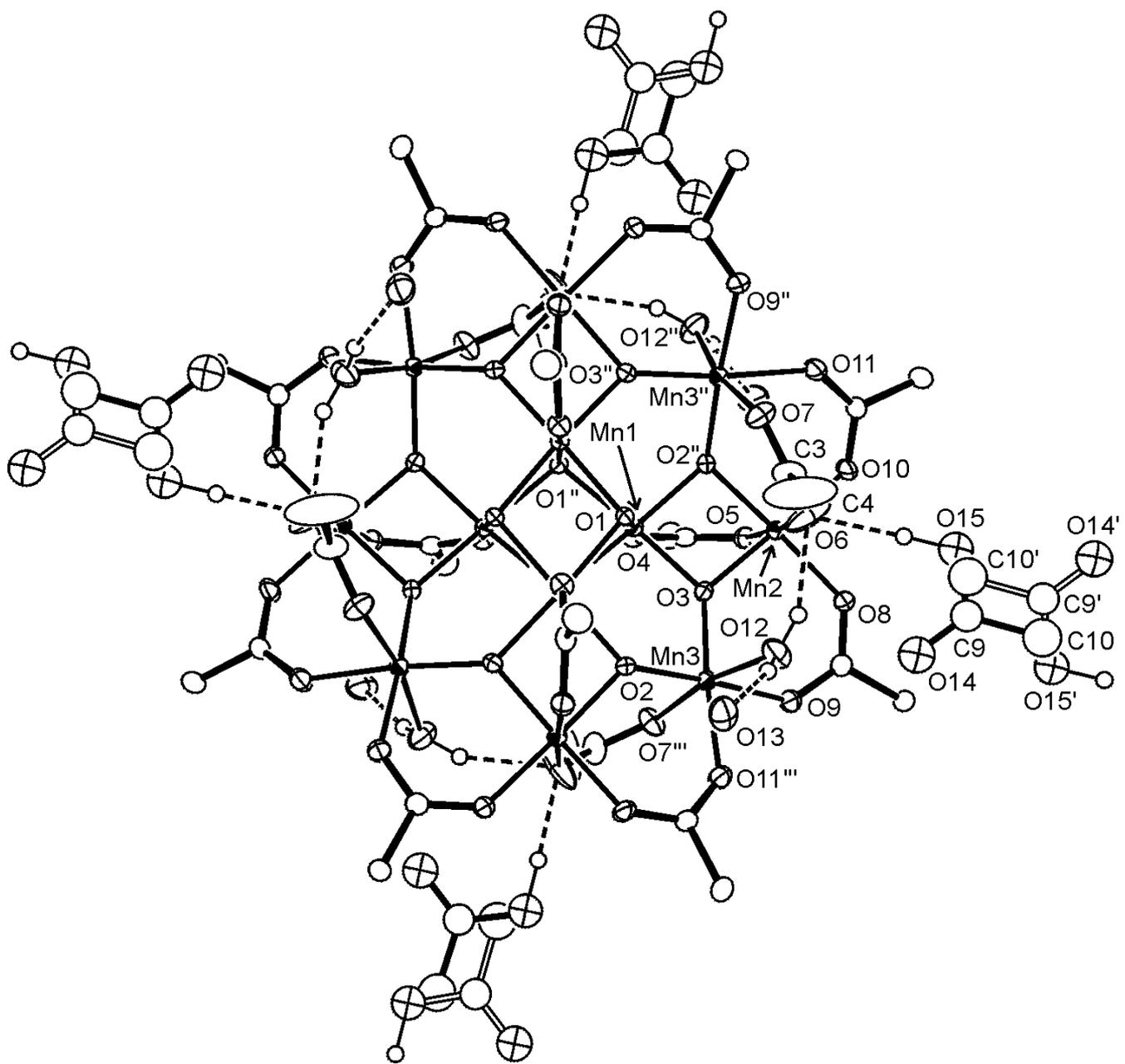

Figure 1

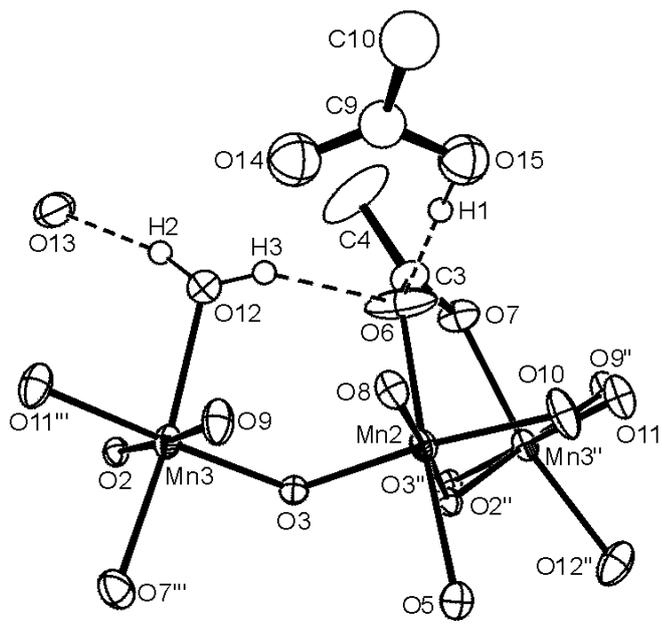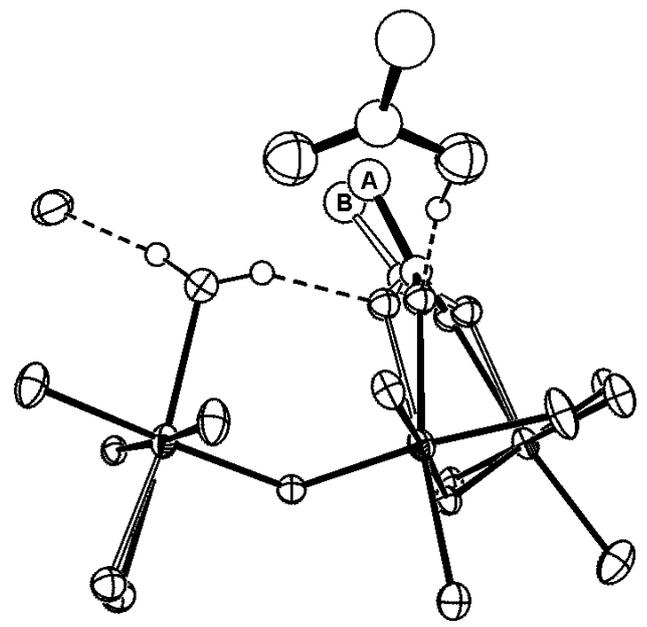

Figure 2

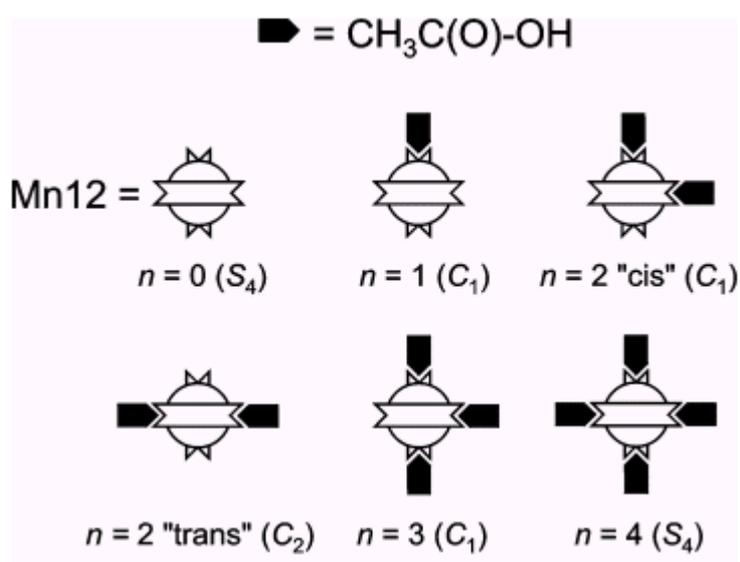

**Figure 3**

**Figure 4.**

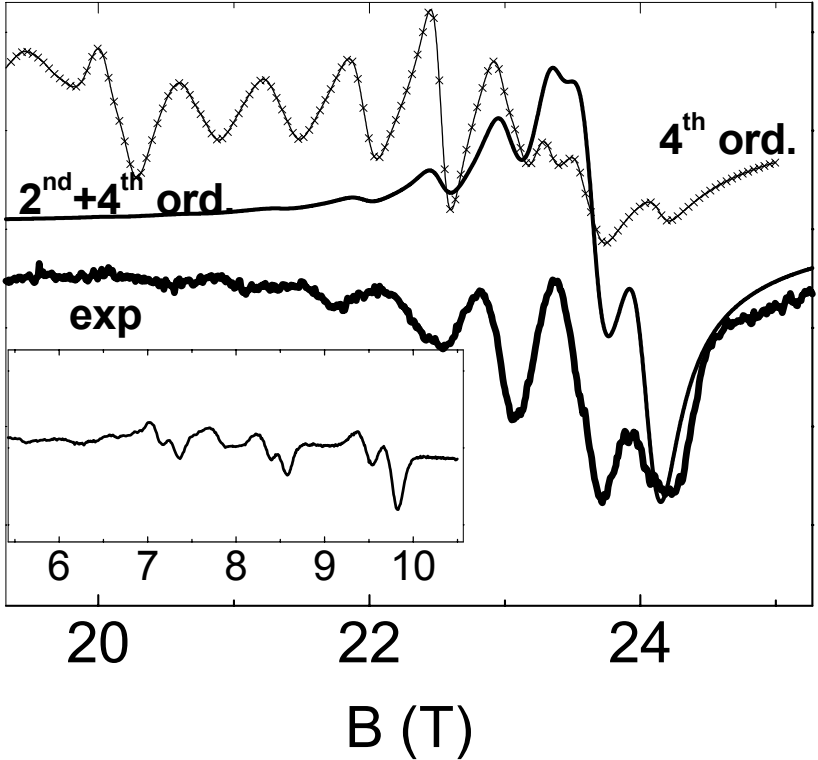